\documentclass[iop]{emulateapj}
\newcommand{\AB}{\ensuremath{0.43^{+0.32}_{- 0.29}}}
\newcommand{\eps}{\ensuremath{1.47^{+0.30}_{- 0.25}}}
\newcommand{\tran}{\ensuremath{1.31^{+ 0.28}_{- 0.13}}}
\newcommand{\Amplitude}{\ensuremath{127^{+12}_{-14}}}
\newcommand{\Depth}{\ensuremath{181^{+15}_{-19}}}
\newcommand{\Offset}{\ensuremath{-34^{+4}_{-5}}}
\newcommand{\Tday}{\ensuremath{2573^{+120}_{-153}}}
\newcommand{\Tmin}{\ensuremath{2709^{+129}_{-159}}}
\newcommand{\Tmax}{\ensuremath{1613^{+118}_{-131}}}
\newcommand{\Mp}{\ensuremath{8.08 \pm 0.31 M_{\Earth}}}
\newcommand{\Rp}{\ensuremath{1.91 \pm 0.08 R_{\Earth}}}

\begin{document}

\title{A Case for an Atmosphere on Super-Earth 55 Cancri e}

%\correspondingauthor{Renyu Hu}
%\email{renyu.hu@jpl.nasa.gov}

\author{Isabel Angelo}
\affiliation{Jet Propulsion Laboratory, California Institute of Technology, 4800 Oak Grove Drive, Pasadena, CA 91109, USA \\
Department of Astronomy, University of California, Berkeley CA, 94720, USA}

\author{Renyu Hu}
\affiliation{Jet Propulsion Laboratory, California Institute of Technology, 4800 Oak Grove Drive, Pasadena, CA 91109, USA \\
Division of Geological and Planetary Sciences, California Institute of Technology, Pasadena, CA 91125, USA}
\email{renyu.hu@jpl.nasa.gov}

\begin{abstract}
%With the number of known exoplanets growing each year, it has become of interest to determine their atmospheric properties and surface conditions. 
One of the primary questions when characterizing Earth-sized and super-Earth-sized exoplanets is whether they have a substantial atmosphere like Earth and Venus or a bare-rock surface like Mercury. Phase curves of the planets in thermal emission provide clues to this question, because a substantial atmosphere would transport heat more efficiently than a bare-rock surface. Analyzing phase curve photometric data around secondary eclipse has previously been used to study energy transport in the atmospheres of hot Jupiters. Here we use phase curve, {\it Spitzer} time-series photometry to study the thermal emission properties of the super-Earth exoplanet 55 Cancri e. We utilize a semi-analytical framework to fit a physical model to the infrared photometric data at 4.5 $\mu$m. The model uses parameters of planetary properties including Bond albedo, heat redistribution efficiency (i.e., ratio between radiative timescale and advective timescale of the atmosphere), and atmospheric greenhouse factor. 
%in addition to determining the reflective and thermal components of the planetary flux.
The phase curve of 55 Cancri e is dominated by thermal emission with an eastward-shifted hot spot. We determine the heat redistribution efficiency to be \eps, which implies that the advective timescale is on the same order as the radiative timescale. This requirement cannot be met by the bare-rock planet scenario because heat transport by currents of molten lava would be too slow. The phase curve thus favors the scenario with a substantial atmosphere. Our constraints on the heat redistribution efficiency translate to an atmospheric pressure of $\sim1.4$ bar. The \textit{Spitzer} 4.5-$\mu$m band is thus a window into the deep atmosphere of the planet 55 Cancri e.
\end{abstract}

\keywords{planets and satellites: atmospheres; planets and satellites: terrestrial planets; planets and satellites: individual (55 Cnc e); occultations; techniques: photometric}

\section{Introduction}\label{sec:intro}

Recent successes in exoplanet science can be attributed in part to the development of high precision photometry. Thousands of known exoplanets were first detected using photometric data from space-based missions like \textit{Kepler}, K2 and \textrm{CoRot}, and this number will continue to grow with the next generation of exoplanet detection missions like TESS \citep{Ricker2016} and PLATO \citep{Rauer2016}. Photometric data are also useful for analysis beyond planet detection. This is because light curves out-of-transit contain light from reflected stellar radiation and planetary thermal emission. Photometric measurements with {\it Spitzer} \citep[e.g.,][]{Knutson2007}, {\it Kepler} \citep[e.g.,][]{Demory2013}, and {\it Hubble} \citep[e.g.,][]{Stevenson2014} have provided insight into the properties of exoplanets' atmospheres. 

Several previous studies have monitored transiting hot Jupiters during and between transit and occultation (or secondary transit), and used phase curves taken in the visible or infrared wavelengths to determine a variety of planet properties \citep[e.g.,][]{Knutson2007,Cowan2012,Knutson2012,Demory2013,Esteves2013,Esteves2015,Stevenson2014,ShporerHu2015,Armstrong2016,Wong2016,Lewis2017}. These phase curves have provided constraints on the temperature of the planet's atmosphere and the longitudinal location of hotspots. In the cases where a reflected stellar radiation component is detected, the phase curves also constrain the location of clouds. These constraints allow us to further study physical properties of the atmosphere such as circulation patterns, temperature profile, and possible molecular composition.

When it comes to Earth-sized and super-Earth-sized exoplanets that may be predominantly rocky, radiation from the planets may come from either the atmosphere or the surface. If the planet has a substantial atmosphere, the phase curve signal would be controlled by temperature and cloud distributions in the atmosphere \citep[e.g.,][]{Hu2015,Webber2015,Parmentier2016}; whereas if the planet has a bare-rock surface, the phase curve would be controlled by the temperature of the surface and patchy surface features, such as lava lakes that affect the reflectivity \citep[e.g.,][]{Kite2016}.

%Phase curves in the visible and infrared can be separated into a variety of components. The first is the light from the host star that is reflected by the planet. This reflected light can then be further split into two sub-components: symmetric reflection (independent of longitude) and asymmetric reflection which may be due to patchy surface features such as clouds or lava lakes. The second component of the phase curve is flux from the planet's thermal emission. This component may be symmetric or asymmetric depending on the existence of surface features than may affect the planet's heat absorption and/or emission (i.e. hotspots, patchy clouds, etc.). 

The first phase curve of a super-Earth exoplanet has been detected recently in the infrared \citep{Demory2016b}. The planet 55 Cancri e has a measured mass of \Mp \ and a radius of \Rp. Interior composition models have found that the measured mass and radius are consistent with a planetary scenario with a massive, high-mean-molecular-weight atmosphere \citep{Demory2011,Winn2011}, or a volatile-poor planetary scenario with a carbon-rich interior and no atmosphere \citep{Madhusudhan2012}. It is also known that the planet does not have an extended, H-rich exosphere \citep{Ehrenreich2012}, which excludes the possibility of an H-rich atmosphere, but does not exclude the high-mean-molecular-weight atmosphere scenario. There have been searches for molecular features of 55 Cancri e, using transit spectroscopy in the infrared \citep{Tsiaras2016} and high-resolution optical spectroscopy \citep{Esteves2017}, but results are non-conclusive. The phase curve of 55 Cancri e, taken in the \textit{Spitzer} Infrared Array Camera (IRAC) 2 band of 4 -- 5 $\mu$m, features a peak of the planet's radiation occurring prior to the occultation, and a large day-night temperature contrast. \citet{Demory2016b} hypothesized that the phase curve could be explained by either a planet with an optically thick, high-mean-molecular-weight atmosphere, or a planet devoid of atmosphere with low-viscosity magma flows.

% Interpretation of this phase curve, however, is far less clear. 

In this paper, we utilize a physical model developed by \citet{Hu2015} to analyze the phase curve for 55 Cancri e. The purpose of our study is to augment the results of \citet{Demory2016b} by using the same data set to fit for a more physically motivated model and derive improved constraints on the planet's heat redistribution. Our re-analysis of the phase curve suggests that the planet has a substantial atmosphere and the \textit{Spitzer} IRAC 2 band is a window into the pressure level as deep as 1 -- 2 bars. We first outline our data preparation and model fitting in \S\ref{sec:analysis}. The results of our analysis are presented in \S\ref{sec:results}. In \S\ref{sec:discussion} we interpret these results and discuss their implications on the planetary scenarios. Finally, we summarize our findings and discuss prospects for future research in \S\ref{sec:conclusion}.

%In this paper, we utilize methods developed and outlined in \citet{Hu2015} in order to determine the specific contributions of each component -- symmetric reflection, asymmetric reflection, and thermal emission -- to the infrared light curve of the super-Earth 55 Cancri e taken by the \textit{Spitzer} spacecraft. This involved fitting a model of the planet's reflection and emission as a function of orbital phase to the photometric light curves before, during and after secondary eclipse. We also derive a set of occultation parameters from our fitted light curve that we use to determine the planet's Bond albedo, heat redistribution efficiency, and greenhouse factor.

%The transit of super-Earth 55 Cancri e was first detected with photometry provided by both \textit{Spitzer} and MOST telescopes \citep{Demory2011, Winn2011}. Since then, there has been ongoing debate over the planet's interior composition and the existence of a water- or volatile-rich envelope \citep{MadhusudhanRedfield2015, Lopez2016,Esteves2017}. While some studies have indicated the presence of a thick and potentially water-bearing atmosphere \citep{Gillon2012,Ridden-Harper2016,Tsiaras2016}, others are consistent with a rocky core without a volatile cover \citep{Ehrenreich2012, Madhusudhan2012,Demory2016b}.

\section{Analysis}\label{sec:analysis}
\subsection{Observations}\label{sec:obs}
For our analysis we use  the same photometric data as \citep{Demory2016b}, taken by \textit{Spitzer} Space Telescope Infrared Camera (IRAC 2) at $4.5 \mu$m. Observations during the primary transit and occultation of 55 Cancri e were taken between 15 June and 15 July 2013. A total of 4,981,760 frames were obtained with  an integration time of 0.02 seconds. The frames were taken during 8  observing sessions of 9 hours each (half of the planet's orbital period), yielding a total observation time of 75 hours.

\subsection{Data Reduction and Preparation}\label{sec:dataprep}

Once the observations were obtained from \textit{Spitzer}, they needed to be converted into a photometric time series from which we can perform phase curve analysis and modeling. The raw photometric time series was computed from the individual frames by \citet{Demory2016b} using methods outlined in \citet{Demory2011}. These raw data were subject to noise from the IRAC detector itself as well as  correlated noise from \textit{Spitzer} as it moves during observation. This noise was removed from our data set by \citet{Demory2016b} to produce a final time series of 30-second bins with an average error that we calculate to be $\sim~360$ parts per million.  

\begin{figure*}
\centering
\includegraphics[width = \textwidth]{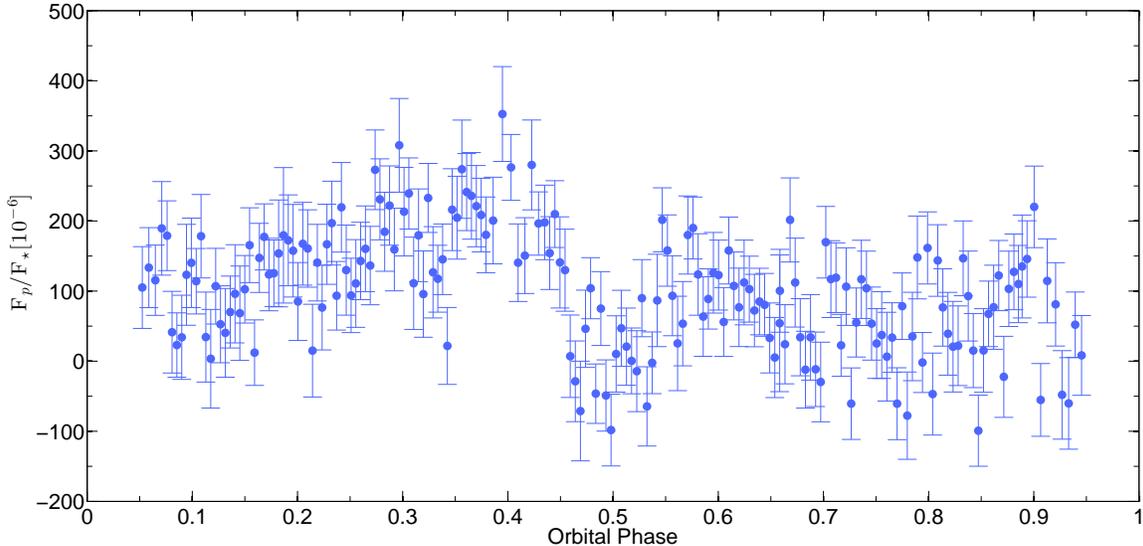}%scale = 0.21 if in single column
\caption{Occultation photometric time series data for 55 Cancri e. Data points represent scaled planet flux $F_p/F_\star$ in parts per million (ppm) and are plotted versus orbital phase, with a phase of 0.5 corresponding to secondary transit. Data are binned per 20 minutes, and each flux value represents the bin sample mean. The uncertainty for each point was calculated by taking the true standard deviation of the bin sample mean $\frac{\sigma}{\sqrt{N}}$.}
\label{fig:phasecurve}
\end{figure*}

After the raw frames were converted into photometric time series data, we then binned the data to ~200 bins for improved visual inspection. The flux value $F_p/F_\star$ relative to the in-occultation stellar flux for each bin was computed by averaging the 30-second flux values for each sample. Uncertainties in each bin were computed by dividing the standard deviation of the sample by the square root of the total sample size per bin, yielding an average bin error of $\sim56$ parts per million. We then phase-folded our time series data and removed data taken during primary transits that are not accounted for in the phase curve model we implemented. While not fitting the primary transits, we use the constraints from the primary transits on planetary radius, semi-major axis, and impact parameter in the subsequent phase curve modeling (see Section \ref{sec:mcmc}). Our photometric time series is shown in Figure \ref{fig:phasecurve}.

Several effects beyond those of planetary atmospheric processes can determine the properties of an occultation phase curve. For example, stellar variability brought about by magnetic field interactions with the planet can cause star spots with a period similar to the planet's orbital period \citep{Shkolnik2008}. Additionally, gravitational interactions between a star and its planetary companion can produce effects such as relativistic beaming and tidal ellipsoidal distortion modulations that can alter the phase curve \citep[e.g.][]{Shporer2017,ShporerHu2015}. These processes do not produce significant effects on the infrared light curve for 55 Cancri e \citep[e.g., beaming modulation $\sim1$ ppm, ellipsoidal modulation $\sim0.6$ ppm;][]{Demory2016b}. We therefore make no corrections to our sample to account for these effects. Furthermore, the orbit of 55 Cancri e can be approximated as circular \citep[$e\approx0.040\pm0.027$, see][]{Baluev2015}, allowing us to use a model fitting method in line with that of \citet{Hu2015}. 

One must be cautious when binning time series photometric data for phase curve analysis to prevent loss or distortion of information \citep[e.g.,][]{Kipping2010}. We have additionally computed model fits for the same data with a much smaller bin size (the native 30-second bins), and with a larger bin size similar to the phase curve in \citet{Demory2016b} (a total of $\approx70$ bins of in- and out-of-transit data). The constraints on the jump parameters are identical to the 200-bin data set to within uncertainty limits, confirming that our model fit does not depend on the binning of the data. We therefore proceed with our analysis using only results from the $\approx200$-bin data set described above.

\subsection{Phase Curve Modeling}\label{sec:mcmc}

To perform our model fit to the photometric data, we use a semi-analytical method outlined in \citet{Hu2015}. The general model performs a fit to the occultation light curve of a planet and also computes three independent light curves representing flux contributions from symmetric reflection, asymmetric reflection (e.g., due to patchy clouds), and thermal emission. It also uses an existing model of secondary transit \citep{MandelAgol2002} from which it can derive the phase curve occultation depth, phase amplitude, and phase offset. 

As shown in \citet{Hu2015}, the scaled total reflective component (i.e. symmetric + asymmetric) of the transiting planet $\frac{F_R}{F_{\star}}$ is computed as follows:
\begin{equation}
\frac{F_R}{F_{\star}} = A_G(\frac{R_p}{a})^2
\end{equation}
where where $A_G$ is the uniform geometric albedo of the planet, $R_p$ is the planet radius, and $a$ is the semi-major axis of the planet's orbit. This means that the contribution of the reflective components to the planet's phase curve is largely governed by the magnitude of $(\frac{R_p}{a})^2$. For 55 Cancri e, using values for $R_p$ and $a$ found in \citet{Demory2016b}, we find that this value is $2.77 \times 10^{-5}$, or $\approx 28$ parts per million (p.p.m.). This value, even with a geometric albedo of 1, would thus contribute no more than 28 p.p.m. to our final model fit which takes on much larger flux fluctuations $\sim 200$ p.p.m. The data, with a dispersion of 56 ppm, are too noisy to detect any reflected light component. It is also not possible to distinguish between uniform and patchy reflective clouds or surfaces, because the asymmetric reflection component would make a contribution much smaller than 28 ppm to the light curve. We therefore assume for the purposes of this paper a purely symmetric reflection component and take the asymmetric component of the reflection contribution to be zero at all phase values.

The general model fit computes the posterior distributions of three to five parameters from the phase curve. The first is the Bond albedo ($A_B$) which is a measurement of the fraction of incident light on the planet that is scattered back into space. The second, referred to as the heat redistribution efficiency ($\epsilon$), is defined as
\begin{equation}
\epsilon = \frac{\tau_{rad}}{\tau_{adv}},
\end{equation}
where $\tau_{rad}$ is the radiative timescales and $\tau_{adv}$ is the advective timescale \citep{Cowan2011,Hu2015}. This parameter describes how well heat is transported from the planet's dayside to nightside. When $|\epsilon|\gg1$ heat transport is more efficient than radiative cooling and the longitudinal variation of temperature will be small, corresponding to giant planets in the Solar System; when $|\epsilon|\ll1$ transport is minimal and the planet will be in local thermal equilibrium, leading to large day-night contrast \citep{Cowan2011}. The sign of $\epsilon$ is the same as the sign of $\tau_{adv}$, which indicates the direction of transport and the thermal phase shift: when $\epsilon > 0$ the transport is eastward and the peak of the thermal emission appears prior to the occultation; when $\epsilon < 0 $ the transport is westward and the peak of the thermal emission appears after the occultation. The magnitude of the radiative timescale depends on the heat capacity and the temperature, and the magnitude of the advective timescale is determined by the speed of fluid movement (atmosphere or molten lava), which is discussed further in Section \ref{sec:discussion}.

The third parameter is the greenhouse factor ($f$), a measurement of the extent to which infrared radiation is absorbed by greenhouse gases in the atmosphere. It is defined as the ratio between the brightness temperature of the photosphere and the equilibrium temperature. The last two parameters are the cloud condensation temperature ($T_c$), under which the atmosphere precipitates and forms patchy clouds, and the reflectivity boosting factor ($\kappa$) which describes the proportional increase in the reflectivity of the clouds in the atmosphere. The first three parameters ($A_B, \epsilon, f$) are computed for each model while the last two ($T_c$ and $\kappa$) are only relevant in the scenario in which the planet has a patchy cloud covering \citep[see][for more details]{Hu2015}. Because we are assuming the surface of the planet to be symmetrically reflective (i.e. without a patchy cloud or lava lake), $T_c$ and $\kappa$ are not used by the phase curve model. We are thus performing a model fit that depends on three free parameters: Bond albedo, heat redistribution efficiency, and greenhouse factor.

The methods developed and outlined in \citet{Hu2015} use a Markov Chain Monte Carlo (MCMC) to compute both the fitted parameters and their posterior distributions. We first input the allowed ranges within which the MCMC can sample for each parameter. The Bond albedo is defined as a fraction that naturally ranges in [0,1]. The heat redistribution $\epsilon$ is allowed to range between -100 and 100, where the sign is determined by the direction of the photosphere rotation with respect to the planet's orbit. For the atmospheric greenhouse factor $f$, we allowed a range of [1,2] with 1 corresponding to no greenhouse effect and 2 corresponds to a thermal photosphere temperature 2 times greater than the planet's equilibrium temperature. Allowing $f > 2$ is not necessary because the temperature of the planet does not have to double even for $A_B$ approaching 1 (see \S\ref{sec:fittedparameters}). These input ranges are wide enough and do not cause distortion in the parameters' posterior distributions; we confirm this by inspecting the output of the Markov chains, which shows the posterior distributions are narrower than the prior ranges (Figures \ref{fig:distributions} and \ref{fig:correlation}).

In addition to the allowed ranges for the fitted parameters, we input values for system parameters that the algorithm uses to compute the model phase curves. We use an empirical and absolutely calibrated infrared spectrum of 55 Cancri \citep{Crossfield2012} to compute the stellar flux in the IRAC 4.5-micron bandpass, and also use the scaled semi-major axis ($a/R_{\star}$), the ratio between the planet's radius and the star's radius ($R_p/R_{\star}$), and the impact parameter ($b$) constrained by primary transit observations \citep{vonBraun2011,Demory2011,Demory2016b}. We use the value of $R_p/R_{\star}$ measured from the same observation as the phase curve \citep{Demory2016a}, which is, $365\pm25$ ppm, corresponding to $R_p/R_{\star}=0.0191\pm0.0007$, in our analysis. The stellar luminosities and orbital parameters are known to high precisions, and they make negligible contributions to the uncertainties of our fitted parameters. The planetary radius is more uncertain and we thus propagate its uncertainty to the fitted parameters in \S~\ref{sec:fittedparameters}.

%although our final model and fitted parameters depend on the values for $a/R_{\star}$ and $R_p/R_{\star}$, it is not necessary propagate their errors for the purposes of this paper. We calculate the errors for these values to be $\lesssim$ 1\%. This fractional error is much smaller than that of the fitted parameters (see \S\ref{sec:fittedparameters}) and therefore make negligible contributions to the uncertainties of our final fitted parameters. 

%Additionally, the star's effective temperature has been previously constrained to within 0.4\% \citep[$T_{eff} = 5196\pm24$, see][]{vonBraun2011} and therefore cannot produce a significant uncertainty in our model phase curve \citep{Hu2015}.

Next we computed two separate Markov chains, each with 100,000 steps representing different combinations of the fitted parameters. The first half of each chain, which we refer to as the ``burn-in" period, were discarded from our final results. We then confirmed the robustness of our final results by verifying their convergence \citep[i.e. assuring $R < 1.01$ for all parameters, as described in ][]{GelmanRubin1992}. 

\section{Results}\label{sec:results}

\subsection{Light Curve Parameters}\label{sec:derivedparameters}

\begin{figure*}
\centering
\includegraphics[width = \textwidth]{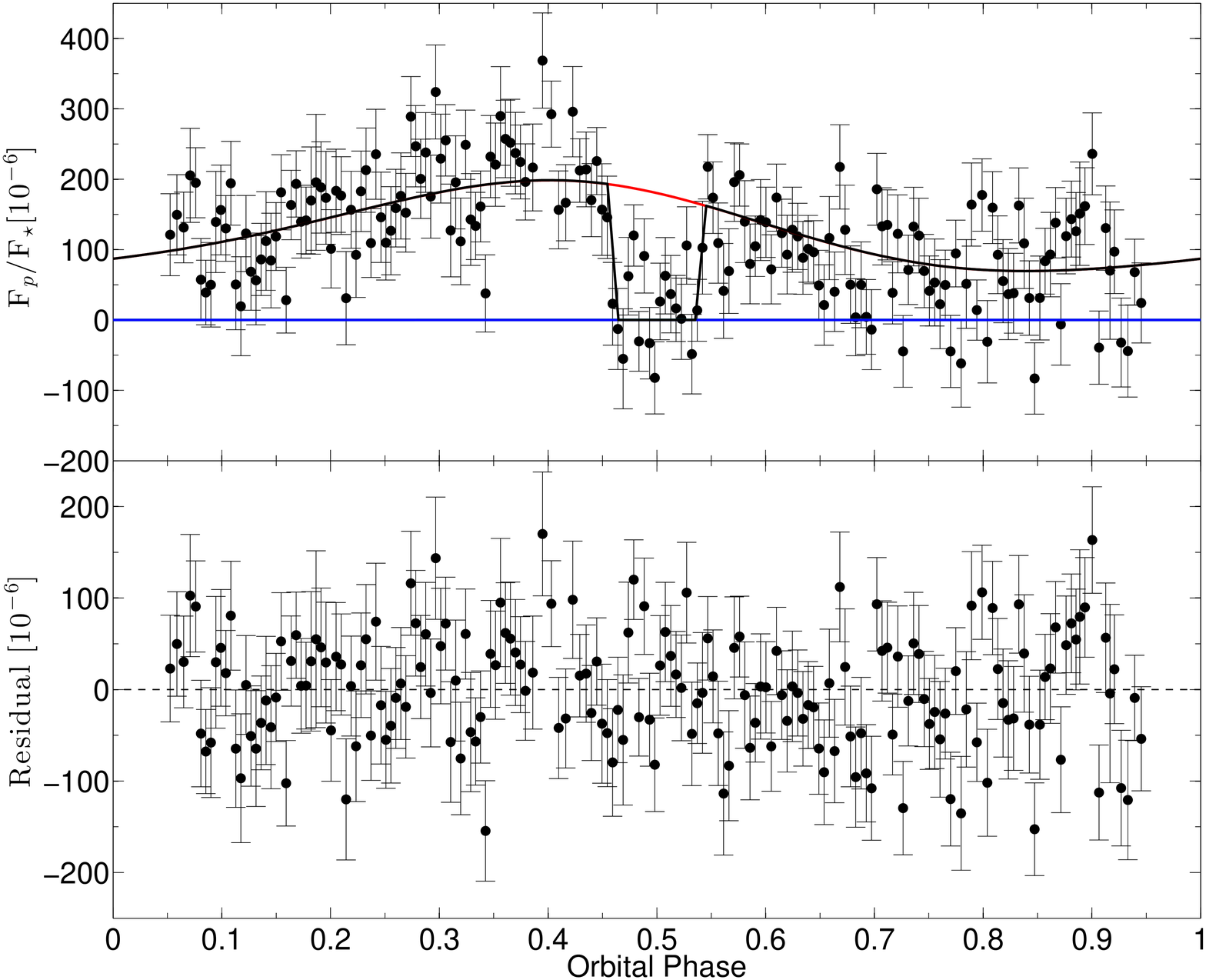}
\caption{The top panel shows the model fit to our phase curve plotted against photometric data from Figure \ref{fig:phasecurve}. The black curve represents the best-fit model phase curve, and the rest show flux contributions from thermal emission (red) and reflection (blue). The model fit is governed primarily by the planet's thermal emission and has an asymmetry due to an eastward-shifted hot spot. In the bottom panel the residuals of the model fit are plotted for each of the photometric data points.}
\label{fig:modelfit}
\end{figure*}

The fitted and derived system parameters for 55 Cancri e are listed in Table \ref{tab:parameters}. The values for these parameters were computed by taking the median of the posterior distribution generated by the MCMC analysis (see \S\ref{sec:mcmc}). The final model fit and residuals computed from these parameters are plotted against our occultation data in Figure \ref{fig:modelfit}. From the model fit we can see that the phase curve of 55 Cancri e is dominated by thermal emission with no significant contribution from the reflected light. We can also see that the fit to the light curve is not symmetric about the secondary transit. Instead, there is a phase curve maximum that occurs pre-occultation at an offset of $\approx-34^{\circ}$.

\begin{deluxetable}{lcc}
\centering
\tabletypesize{\scriptsize}
\tablewidth{0pc}
%\tablenum{1}
\tablecaption{Fitted and Derived Parameters for 55 Cancri e$^\textrm{a}$}
\tablehead{\colhead{Parameter} & \colhead{Value}}
\startdata
\sidehead{\textit{Derived Parameters}}
Phase amplitude (ppm) & \Amplitude \\
Eclipse depth (ppm) & \Depth \\
Phase offset (degree)$^\textrm{b}$ & \Offset \\
\sidehead{\textit{Fitted Parameters}}
Bond albedo $A_B$& \AB \\
Heat redistribution $\epsilon$ $^\textrm{c}$ & \eps\\
Greenhouse factor $f$& \tran \\
\sidehead{\textit{Calculated Surface Temperatures}}
Maximum hemisphere-averaged temperature (K) & \Tmin\\
Minimum hemisphere-averaged temperature (K) & \Tmax\\
Average dayside temperature (K) & \Tday\\
\enddata
\tablecomments{\\
$^\textrm{a}$ The best fit model has $\chi^2/\textrm{dof of }     1.14$\\
$^\textrm{b}$ Defined to be positive for post-occultation maximum\\
$^\textrm{c}$ Defined to be positive for eastward-traveling winds in a synchronously rotating reference frame\\}
%$^\textrm{b}$ Longitudinal boundary of clouds in the western hemisphere, with the substellar point defined as zero longitude\\
%$^\textrm{c}$ Longitudinal boundary of clouds in the easter hemisphere, where 90$^{\circ}$ signifies that the eastern hemisphere is cloud-free\\
%$^\textrm{d}$ Reflectivity of the clear longitudes \\
%$^\textrm{e}$ Reflectivity of the cloudy longitudes}
\label{tab:parameters}
\end{deluxetable}

The phase curve model from \citet{Hu2015} generates the amplitude, occultation depth and offset based on the fit to the observed data (see Table \ref{tab:parameters}). Errors for all values in Table \ref{tab:parameters} were computed using the values 1 standard deviation above and below the mean, corresponding to z scores of $\pm1$ for a standard normal distribution. Here we measure our phase-curve amplitude to be \Amplitude\ p.p.m. This value is $\approx 2\sigma$ smaller but better constrained than the amplitude found in \citet{Demory2016b}. For our eclipse depth we derive a value of \Depth\ p.p.m., which is $1\sigma$ greater than the value derived in \citet{Demory2016b}. Upon inspection of the model phase curves in \citet{Demory2016b}, the best-fit models look similar at the peak ($\frac{F_p}{F_{\star}}$ around $\sim 200$ p.p.m.), but our best-fit model indicates a minimum out-of-transit value of $\sim 70$ p.p.m., as opposed to $\sim 50$ p.p.m. of \citet{Demory2016b}. The difference is mostly due to the different models used in the phase curve analysis. \citet{Demory2016b} used a single-longitude-band model and a three-longitude-band model to fit the phase curve, and their models have sharp temperature discontinuities between the bands. Our model does not allow for these discontinuities and instead calculates a smooth longitudinal distribution of temperature by solving an energy transport equation \citep[see][]{Hu2015}. Our physically motivated model provides a good fit to the phase curve, and indicates a smaller phase amplitude and a higher nightside temperature than \citet{Demory2016b}.
%Our eclipse depth reported here is larger than the phase amplitude by almost $4\sigma$, indicating a clear detection of the nightside emission. 
This analysis thus highlights the need to use a physically motivated model in the phase curve analysis compared to the longitudinal band models.

%However, upon inspection of the phase curve in \citet{Demory2016b}, our models look similar- both have a peak $\frac{F_p}{F_{\star}}$ around $\sim 200$ p.p.m. and a minimum out-of-transit value of $\sim 150$ p.p.m. We therefore require further analysis of the two model fits to determine the source of discrepancy for the two computed phase amplitude values.}

Our third derived parameter is the phase curve offset, defined to be positive when the phase curve maximum occurs post-occultation. We find a pre-occultation phase curve offset of \Offset\ degrees which agrees with the offset from \citet{Demory2016b} to within error estimates. Because phase curve is dominated by contributions from thermal emission (Figure \ref{fig:modelfit}), the offset is likely due to a hot spot located east of the substellar point (see \S\ref{sec:discussion} for more details).

The posterior probability distributions for our fitted model parameters are shown in the top 3 panels of Figure \ref{fig:distributions}. As we can see, the three derived parameters discussed above are tightly constrained in the scenario to which we fit our model curve. In other words, 
%in the case that the planetary light from 55 Cancri e has negligible contributions from asymmetrically reflective clouds or lakes, 
the phase curve amplitude, offset, and occultation depth are well constrained to the values listed in Table \ref{tab:parameters}.

We also used our derived phase curve parameters to compute a series of surface temperatures for the planet (Table \ref{tab:parameters}). We used the eclipse depth to compute an average dayside temperature of \Tday\ K. From the phase curve amplitude we calculate a minimum and maximum hemisphere-averaged temperature of \Tmin\ K and \Tmax\ K respectively. All of these values agree with those found in \citet{Demory2016b} to within 1$\sigma$, but we compute a higher nightside temperature and a smaller day-night temperature contrast of $\sim 950$K that is more in-line with the presence of a convective envelope. The fact that our maximum hemisphere-average temperature is greater than the average dayside temperature is consistent with our phase curve offset of \Offset\ and is likely due to an eastward-shifted hotspot. These implications are discussed further in \S\ref{sec:discussion}.

\subsection{Model Parameters}\label{sec:fittedparameters}

In addition to the derived parameters discussed in \S\ref{sec:derivedparameters}, the model phase curve contains enough information to compute a set of fitted planetary parameters (see \S\ref{sec:mcmc}). In the case of a homogeneous reflective layer, the model outputs estimates for the planet's Bond albedo $A_B$, heat redistribution efficiency $\epsilon$, and greenhouse factor $f$. The fitted values for these parameters can be found in Table \ref{tab:parameters}.

\begin{figure*}
\centering
\includegraphics[width = \textwidth]{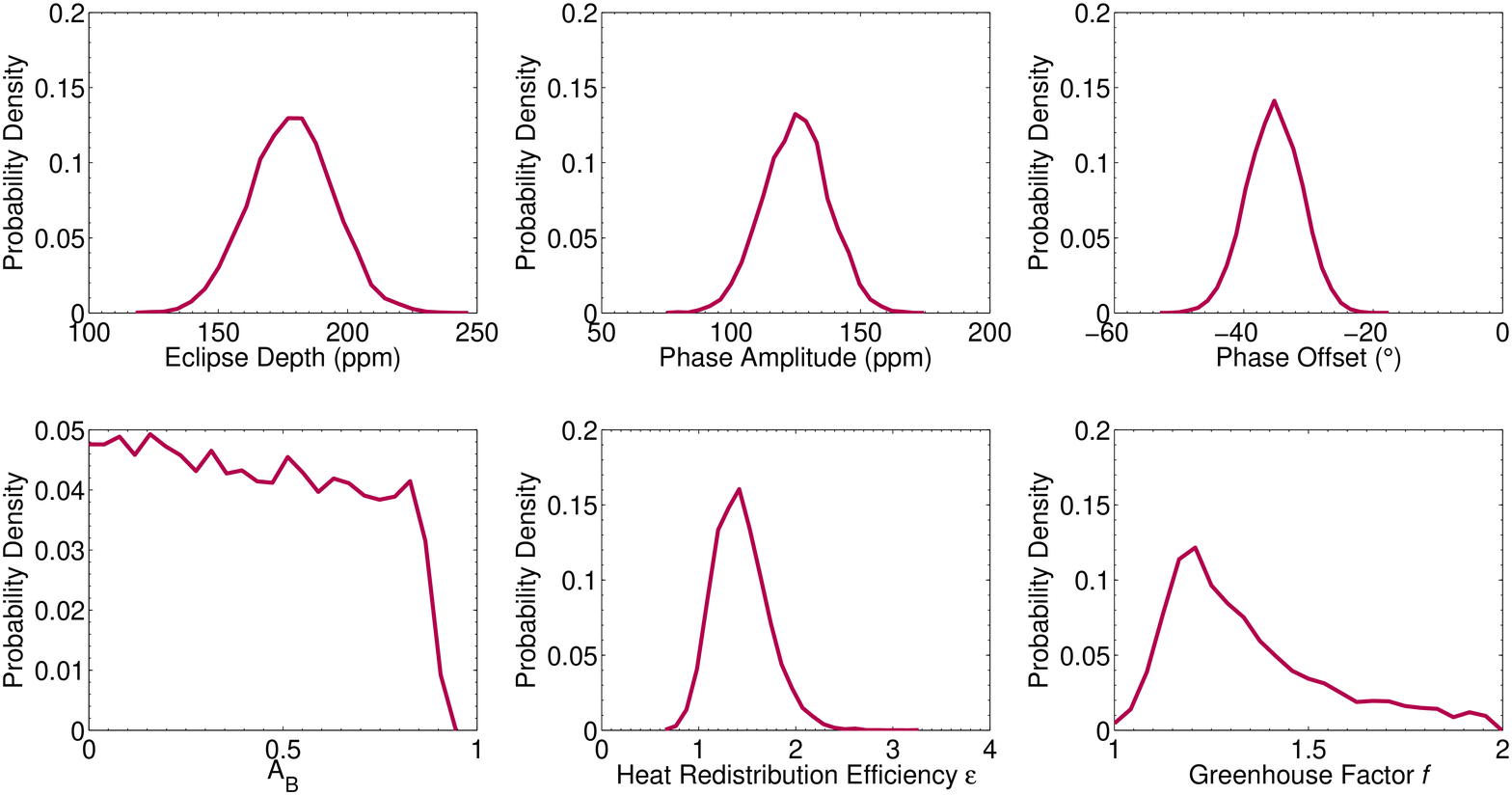}%scale = 0.21 if in single column
\caption{Posterior probability distributions for the derived (top) and fitted (bottom) parameters from our MCMC phase curve analysis are plotted for 55 Cancri e. Of the derived parameters, the phase offset of \Offset (top right) is best constrained to within 13\% while the eclipse depth and phase amplitude (top left and middle) are constrained to within $\approx$ 10\% their estimated values. For the fitted parameters, the heat redistribution efficiency $\epsilon = \eps$ has a relatively normal spread ranging in $\approx 20\%$ while the bond albedo and greenhouse factor have less regular distributions.}
\label{fig:distributions}
\end{figure*}

The posterior probability distributions of the fitted planetary parameters for 55 Cancri e can be found in the 3 bottom panels of Figure \ref{fig:distributions}. As we can see, the value for the heat redistribution efficiency (middle panel) is well constrained between 1 and 3 at an estimated value of $\epsilon = $\eps. A positive value for $\epsilon$ at almost $5\sigma$ above zero indicates that the advective frequency of the planet's envelope is non-zero and consequently that material at photospheric pressures above the planet's surface travel eastward in a synchronously rotating frame. The magnitude of $\epsilon$ is inconsistent with a lava ocean and thus suggestive of a thick atmosphere. This scenario is discussed in more detail in \S\ref{sec:discussion}.

\begin{figure}
\centering
\includegraphics[width = 0.45\textwidth]{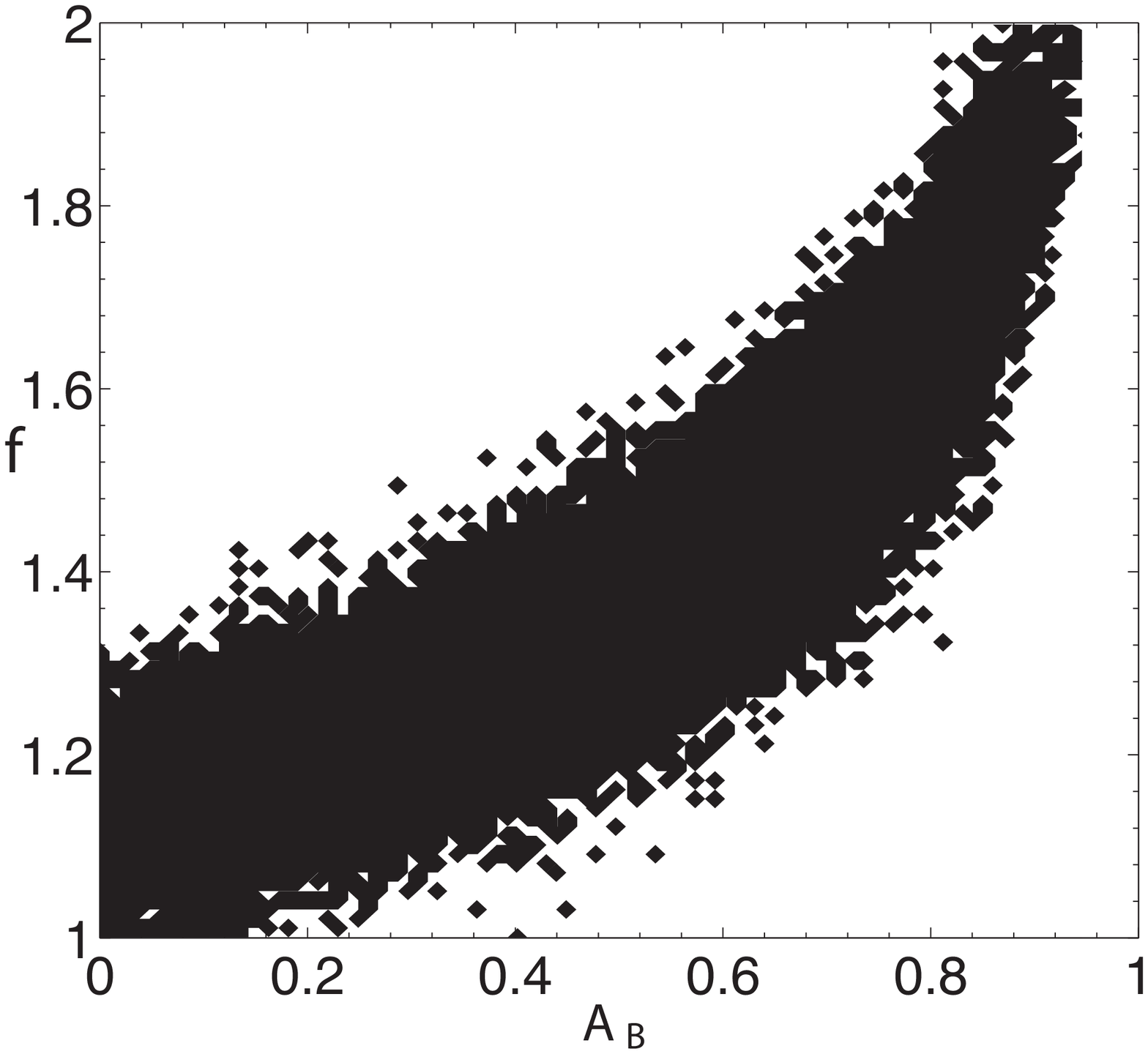}
\caption{Correlation between Bond albedo $A_B$ and greenhouse factor $f$ shows that larger greenhouse effect corresponds to a larger overall planet reflectivity. We see that smaller $A_B \approx 0.5$ corresponds to little to no greenhouse affect at $f \approx 1$. Additionally $f$ quickly approaches 2 as $A_B$ gets closer to 1, a non-physical case for planet atmospheres.}
\label{fig:correlation}
\end{figure}

As we can tell by inspection of Figure \ref{fig:distributions}, the Bond albedo and greenhouse factor do not have the same narrow distribution as $\epsilon$, and they are not constrained independently. The planet's Bond albedo $A_B$ can take any value smaller than 0.9. Similarly, the atmospheric greenhouse factor $f$ is typically closer to 1.2 but also takes essentially all of the allowed values between 1 and 2. The two parameters are correlated, as shown in Figure \ref{fig:correlation}. This is due to the proportionality from \citet{Hu2015}, viz.
\begin{equation}
T \propto f(1-A_B)^{1/4}
\end{equation}
which our model uses to compute the temperature distribution on the planet's surface. As we can see from Equation (2), raising $f$ requires an additional increase in $A_B$ for a given temperature and vice versa. This is in line with what we see in Figure \ref{fig:correlation} where a lower $A_B \approx 0.1$ corresponds to lower $f \leq 1.3$ and $f$ slowly leveling off at 2 as $A_B$ approaches 1. Additionally the fact that $f >2$ corresponds to a forbidden Bond albedo greater than 1 further motivates our allowed ranges for $f$ in our model fit of [1,2]. 

The uncertainty in the planetary radius (8\% in terms of $(R_p/R_{\star})^2$) has an impact on the derived temperatures, and thus the greenhouse factor $f$. We propagate this uncertainty by calculating the brightness temperatures corresponding to the 1$\sigma$ upper and the lower bounds of the planetary radius, and find that this would increase the uncertainty in the dayside and nightside temperatures by 50 K and 20 K respectively. This in turn causes the uncertainty in the greenhouse factor to increase by 0.04, in addition to 0.28 as tabulated in Table \ref{tab:parameters}. The uncertainty in the planetary radius thus has no impact on the general results.

\section{Discussion}\label{sec:discussion}

The model fit to our \textit{Spitzer} light curve for 55 Cancri e is composed primarily of planetary thermal emission with a negligible contribution from reflected light of the host star. The fitted light curve also shows a pre-occultation phase shift of \Offset $^\circ$. Because the planetary light is dominated by thermal emission, we interpret this shift to be due to an eastward-shifted hotspot on the planet surface. This in conjunction with our positive fitted value for heat redistribution efficiency $\epsilon = \eps$ and our divergent values for dayside and maximum temperatures suggest eastward heat transport via advection in the planet's outer envelope.

Our fitted value for $\epsilon$ indicates that the advective timescale is on the same order as the radiative timescale, and implies a thick atmosphere around 55 Cancri e. There are two possible planetary scenarios that determine the value of $\tau_{adv}$. The first is the scenario in which heat is primarily transported by currents in magma lakes. \citet{Kite2016} has estimated the speed of the magma currents driven by the day-night temperature contrast on hot, rocky exoplanets. Neglecting viscosity and inertia, the speed is controlled by the balance between the pressure gradient and the rotational force (i.e., geostrophic balance). The speed is found to be low ($\sim0.02$ m s$^{-1}$), requiring $\sim10$ years for heat to be transport from the dayside to the nightside. The slow transport is due to the very small expansivity of the magma, and then the very small density contrast, compared with an atmosphere. As such, this estimate is independent from the composition of the magma. With our $\epsilon$ value in this case, the radiative timescale would need to be on the order of tens of years, far too long given a dayside temperature $>2000$ K (Table \ref{tab:parameters}). The phase curve thus disfavors the scenario in which heat on 55 Cancri e is transported primarily via lava lakes.

The second possible scenario is that heat is transported convection in a thick atmospheric envelope. In this case, the advective timescale $\tau_{adv} = R_p/v_w$. Using a planet radius of $R_p = 1.91 R_\oplus$ and wind speed of 1,000 m s$^{-1}$ typically obtained in GCMs for high-mean-molecular-weight atmospheres \citep[e.g.,][]{ZhangShowman2017}, we compute an advective timescale of $\tau_{adv} \sim 3$ hours, suggesting that the radiative timescale would need to be on the order of hours, a much more plausible scenario than that of heat transport by lava lakes. 

The radiative timescale of an atmosphere is related to the photospheric-level pressure $P$ as
\begin{equation}
\tau_{rad} = \epsilon \times \tau_{adv} = \frac{c_pP}{g\sigma T_0^3},
\end{equation}
where $c_p$ is the specific heat capacity, $g$ is the surface gravity, $\sigma$ is the Stefan-Boltzmann constant, and $T_0$ is the temperature of radiation for which we approximate as the average dayside temperature (Table \ref{tab:parameters}). We deduce from Equation (4) that the photosphere would need to be at a pressure of $P \approx 1.4$ bar based on the measured value of $\epsilon$.

Is this photospheric pressure reasonable for the {\it Spitzer} observation in 4 -- 5 $\mu$m? Table \ref{tab:opa} lists several hypothetical atmospheric scenarios and their opacities based on simple C-H-O-N-S molecules. The optical depth at 1.4 bar needs to be on the order of, or smaller than, unity in order to be consistent with the photospheric pressure. We see that H$_2$O- or CO$_2$-dominated atmospheres are not possible because of their strong absorption in this wavelength range. However, a CO- or N$_2$-dominated atmosphere is possible. If the atmosphere is made of N$_2$, it should also contain $\sim0.2\%$ of H$_2$O or $2\%$ of CO$_2$ to make the photosphere to be at 1.4 bar. Such abundances of H$_2$O or CO$_2$ seem reasonable for an N$_2$-dominated atmosphere. We also note that an evaporation atmosphere in vapor equilibrium with the magma (made of Na, O$_2$, SiO, Mg, and Fe) would have a total surface pressure of $10^{-3}\sim10^{-2}$ bar depending on the composition of the underlying magma \citep{Schaefer2009}. Such a tenuous vapor atmosphere cannot transport heat fast enough \citep[e.g.,][]{Kite2016} and thus cannot explain the phase curve features.

\begin{table}[htp]
\caption{Opacities of the hypothetical atmospheric scenarios of 55 Cancri e.}
\begin{tabular}{llc}
\hline
Type of Atmosphere & Source of Opacity & $\tau$ at 1.4 bar\\
\hline
%H$_2$-dominated & H$_2$-H$_2$ CIA & 4.3 \\
H$_2$O-dominated & H$_2$O & 450\\
N$_2$-dominated & N$_2$-N$_2$ CIA & 0.06 \\
CO$_2$-dominated & CO$_2$ & 38 \\
CO-dominated & CO & 1.2 \\
O$_2$-dominated & O$_3$ & 0.8 \\
\hline
\end{tabular}
\tablecomments{The optical depth is the column-integrated opacity at 1.4 bar, averaged over the {\it Spitzer} IRAC 4.5-$\mu$m bandpass. The opacities are adopted from the HITRAN 2012 and HITEMP 2010 databases \citep{Rothman2010, Rothman2012}. The temperature is assumed to be 2000 K for simplicity. CIA stands for collision-induced absorption. O$_3$ is assumed to be photochemically produced and is $10^{-3}$ of the abundance of O$_2$ \citep[e.g.,][]{Hu2012}.}
\label{tab:opa}
\end{table}%

Therefore, the phase curve implies that at \textit{Spitzer} wavelengths around 4.5 $\mu$m we are able to see deep into the planet atmosphere. The atmosphere is clear in this band and cannot have dominated abundances of CO$_2$ or H$_2$O that have strong absorption at the observed wavelengths (Table \ref{tab:opa}, also see \citet{HuSeager2014}). It remains possible that the planet is engulfed by a CO- or N$_2$-dominated envelope. Our analysis supports the case of an atmosphere around 55 Cancri e.

One might ask whether a CO- or N$_2$-dominated envelope on the planet 55 Cancri e is stable against intense irradiation that would drive atmospheric escape. Assuming energy-limited escape, \citet{Demory2016b} estimated that a total of 31 kbar worth of atmosphere would have been lost due to stellar irradiation if the planet has a substantial atmosphere. We stress, however, that this 31 kbar corresponds to only 0.5\% of the mass of the planet, while it is not uncommon to find volatile content greater than 2\% in building blocks of terrestrial planets \citep[e.g.,][]{Elkins2008}. The total escape was also likely an overestimate, because the energy-limited escape formula does not apply to the transonic escape regime at high irradiation, as is the case for 55 Cancri e \citep{Johnson2013}. In this regime, the escape flow has high kinetic energy, but the escape rate is low. Therefore, we suggest a CO- or N$_2$-dominated envelope as a plausible scenario for the planet 55 Cancri e.

%In the case that the planet does possess a water ocean, it is likely that the envelope is subject to high photodissociation rates that keep the planet from harboring a detectable H corona. Ly$\alpha$ observations of the host star 55 Cancri during the planet's transit do not show evidence of an upper HI atmosphere, placing an upper limit of $\lesssim 3 \times 10^8 $g s$^{-1}$ on the planet's atmospheric mass loss rate \citep{Ehrenreich2012}. This mass loss rate could not strip the planet of water in the host star's lifetime and is therefore consistent with a massive, photodissociating water envelope. The fact that this envelope does not produce a significant HI-dominant exosphere may be due to the high radiation that the planet receives at close-in separations, preventing the dissociating Hydrogen from accumulating around the planet.

\section{Conclusion}\label{sec:conclusion}

We present analysis of the infrared \textit{Spitzer} phase curve for the super-Earth 55 Cancri e. We fit a theoretical light curve to the photometric time series taken at 4.5$\mu$m using physical models developed in \citet{Hu2015}. 
%Our model computes both the thermal and reflective components of the planetary light curve in addition to providing estimates and distributions for various atmospheric parameters.
With physically motivated models, our analysis of the phase curve of 55 Cancri e confirms and further constrains the eclipse depth and phase offset reported by \citet{Demory2016b} and estimates a smaller phase amplitude and higher nightside temperature than previously found. 

The \textit{Spitzer} phase curve for 55 Cancri e is dominated by thermal emission with a pre-occultation shift due to a hot spot located east of the substellar point.
%Additionally, the constraints that our photometric data place on both derived and fitted parameters suggest that the near- and mid-infrared light \textit{Spitzer} phase curve for 55 Cancri e is dominated by thermal emission with a pre-occultation shift due to a hot spot located east of the substellar point. 
The planet's heat redistribution efficiency constrained by the phase curve $\epsilon = \eps$ requires that 55 Cancri e is shrouded by a thick atmosphere that acts as a primary source of heat transport. This value corresponds to a photospheric pressure of 1.4 bar, consistent with a CO- or N$_2$-dominated atmosphere, with minor abundances of H$_2$O or CO$_2$. 

The \textit{Spitzer} light curve used in both our analysis and \citet{Demory2016b} represent the first phase curve of a super-Earth-sized exoplanet, reflecting recent progress in observing technology and methods. Our analysis of 55 Cancri e will be the first of many, and further observations of super-Earths in the coming years will set up a context in which our results can be further interpreted. Additionally, photometric observations of 55 Cancri e in other wavelengths may shed light and place further constraints on the planetary characteristics found in this paper. We therefore are optimistic that further research will illuminate the nature of 55 Cancri e's atmosphere and super-Earths alike.

\hfill\\
\hfill\\
We thank Brice-Olivier Demory for the reduced 55 Cancri e photometric data. This work uses observations taken by the \textit{Spitzer} telescope operated by staff at the Jet Propulsion Laboratory, California Institute of Technology under a contract with NASA, where this research was carried out. Support for IA's work was provided by the Caltech Summer Undergraduate Research Fellowship (SURF) Program.


\begin{thebibliography}{}
\expandafter\ifx\csname natexlab\endcsname\relax\def\natexlab#1{#1}\fi
\providecommand{\url}[1]{\href{#1}{#1}}

\bibitem[{{Armstrong} {et~al.}(2016){Armstrong}, {de Mooij}, {Barstow},
  {Osborn}, {Blake}, \& {Saniee}}]{Armstrong2016}
{Armstrong}, D.~J., {de Mooij}, E., {Barstow}, J., {et~al.} 2016, Nature
  Astronomy, 1, 0004

\bibitem[{{Baluev}(2015)}]{Baluev2015}
{Baluev}, R.~V. 2015, \mnras, 446, 1493

\bibitem[{{Cowan} \& {Agol}(2011)}]{Cowan2011}
{Cowan}, N.~B., \& {Agol}, E. 2011, \apj, 726, 82

\bibitem[{{Cowan} {et~al.}(2012){Cowan}, {Machalek}, {Croll}, {Shekhtman},
  {Burrows}, {Deming}, {Greene}, \& {Hora}}]{Cowan2012}
{Cowan}, N.~B., {Machalek}, P., {Croll}, B., {et~al.} 2012, \apj, 747, 82

\bibitem[{{Crossfield}(2012)}]{Crossfield2012}
{Crossfield}, I.~J.~M. 2012, \aap, 545, A97

\bibitem[{{Demory} {et~al.}(2016{\natexlab{a}}){Demory}, {Gillon},
  {Madhusudhan}, \& {Queloz}}]{Demory2016a}
{Demory}, B.-O., {Gillon}, M., {Madhusudhan}, N., \& {Queloz}, D.
  2016{\natexlab{a}}, \mnras, 455, 2018

\bibitem[{{Demory} {et~al.}(2011){Demory}, {Gillon}, {Deming}, {Valencia},
  {Seager}, {Benneke}, {Lovis}, {Cubillos}, {Harrington}, {Stevenson}, {Mayor},
  {Pepe}, {Queloz}, {S{\'e}gransan}, \& {Udry}}]{Demory2011}
{Demory}, B.-O., {Gillon}, M., {Deming}, D., {et~al.} 2011, \aap, 533, A114

\bibitem[{{Demory} {et~al.}(2013){Demory}, {de Wit}, {Lewis}, {Fortney},
  {Zsom}, {Seager}, {Knutson}, {Heng}, {Madhusudhan}, {Gillon}, {Barclay},
  {Desert}, {Parmentier}, \& {Cowan}}]{Demory2013}
{Demory}, B.-O., {de Wit}, J., {Lewis}, N., {et~al.} 2013, \apjl, 776, L25

\bibitem[{{Demory} {et~al.}(2016{\natexlab{b}}){Demory}, {Gillon}, {de Wit},
  {Madhusudhan}, {Bolmont}, {Heng}, {Kataria}, {Lewis}, {Hu}, {Krick},
  {Stamenkovi{\'c}}, {Benneke}, {Kane}, \& {Queloz}}]{Demory2016b}
{Demory}, B.-O., {Gillon}, M., {de Wit}, J., {et~al.} 2016{\natexlab{b}}, \nat,
  532, 207

\bibitem[{{Ehrenreich} {et~al.}(2012){Ehrenreich}, {Bourrier}, {Bonfils},
  {Lecavelier des Etangs}, {H{\'e}brard}, {Sing}, {Wheatley}, {Vidal-Madjar},
  {Delfosse}, {Udry}, {Forveille}, \& {Moutou}}]{Ehrenreich2012}
{Ehrenreich}, D., {Bourrier}, V., {Bonfils}, X., {et~al.} 2012, \aap, 547, A18

\bibitem[{Elkins-Tanton \& Seager(2008)}]{Elkins2008}
Elkins-Tanton, L.~T., \& Seager, S. 2008, The Astrophysical Journal, 685, 1237

\bibitem[{{Esteves} {et~al.}(2013){Esteves}, {De Mooij}, \&
  {Jayawardhana}}]{Esteves2013}
{Esteves}, L.~J., {De Mooij}, E.~J.~W., \& {Jayawardhana}, R. 2013, \apj, 772,
  51

\bibitem[{{Esteves} {et~al.}(2015){Esteves}, {De Mooij}, \&
  {Jayawardhana}}]{Esteves2015}
---. 2015, \apj, 804, 150

\bibitem[{{Esteves} {et~al.}(2017){Esteves}, {de Mooij}, {Jayawardhana},
  {Watson}, \& {de Kok}}]{Esteves2017}
{Esteves}, L.~J., {de Mooij}, E.~J.~W., {Jayawardhana}, R., {Watson}, C., \&
  {de Kok}, R. 2017, \aj, 153, 268

\bibitem[{{Gelman} \& {Rubin}(1992)}]{GelmanRubin1992}
{Gelman}, A., \& {Rubin}, D.~B. 1992, Statistical Science, 7, 457

\bibitem[{{Hu} {et~al.}(2015){Hu}, {Demory}, {Seager}, {Lewis}, \&
  {Showman}}]{Hu2015}
{Hu}, R., {Demory}, B.-O., {Seager}, S., {Lewis}, N., \& {Showman}, A.~P. 2015,
  \apj, 802, 51

\bibitem[{{Hu} \& {Seager}(2014)}]{HuSeager2014}
{Hu}, R., \& {Seager}, S. 2014, \apj, 784, 63

\bibitem[{Hu {et~al.}(2012)Hu, Seager, \& Bains}]{Hu2012}
Hu, R., Seager, S., \& Bains, W. 2012, The Astrophysical Journal, 761, 166

\bibitem[{Johnson {et~al.}(2013)Johnson, Volkov, \& Erwin}]{Johnson2013}
Johnson, R.~E., Volkov, A.~N., \& Erwin, J.~T. 2013, The Astrophysical Journal
  Letters, 768, L4

\bibitem[{Kipping(2010)}]{Kipping2010}
Kipping, D.~M. 2010, Monthly Notices of the Royal Astronomical Society, 408,
  1758

\bibitem[{{Kite} {et~al.}(2016){Kite}, {Fegley}, {Schaefer}, \&
  {Gaidos}}]{Kite2016}
{Kite}, E.~S., {Fegley}, Jr., B., {Schaefer}, L., \& {Gaidos}, E. 2016, \apj,
  828, 80

\bibitem[{{Knutson} {et~al.}(2007){Knutson}, {Charbonneau}, {Allen}, {Fortney},
  {Agol}, {Cowan}, {Showman}, {Cooper}, \& {Megeath}}]{Knutson2007}
{Knutson}, H.~A., {Charbonneau}, D., {Allen}, L.~E., {et~al.} 2007, \nat, 447,
  183

\bibitem[{{Knutson} {et~al.}(2012){Knutson}, {Lewis}, {Fortney}, {Burrows},
  {Showman}, {Cowan}, {Agol}, {Aigrain}, {Charbonneau}, {Deming}, {D{\'e}sert},
  {Henry}, {Langton}, \& {Laughlin}}]{Knutson2012}
{Knutson}, H.~A., {Lewis}, N., {Fortney}, J.~J., {et~al.} 2012, \apj, 754, 22

\bibitem[{{Lewis} {et~al.}(2017){Lewis}, {Parmentier}, {Kataria}, {de Wit},
  {Showman}, {Fortney}, \& {Marley}}]{Lewis2017}
{Lewis}, N.~K., {Parmentier}, V., {Kataria}, T., {et~al.} 2017, ArXiv e-prints,
  arXiv:1706.00466

\bibitem[{{Madhusudhan} {et~al.}(2012){Madhusudhan}, {Lee}, \&
  {Mousis}}]{Madhusudhan2012}
{Madhusudhan}, N., {Lee}, K.~K.~M., \& {Mousis}, O. 2012, \apjl, 759, L40

\bibitem[{{Mandel} \& {Agol}(2002)}]{MandelAgol2002}
{Mandel}, K., \& {Agol}, E. 2002, \apjl, 580, L171

\bibitem[{{Parmentier} {et~al.}(2016){Parmentier}, {Fortney}, {Showman},
  {Morley}, \& {Marley}}]{Parmentier2016}
{Parmentier}, V., {Fortney}, J.~J., {Showman}, A.~P., {Morley}, C., \&
  {Marley}, M.~S. 2016, \apj, 828, 22

\bibitem[{{Rauer} {et~al.}(2016){Rauer}, {Aerts}, {Cabrera}, \& {PLATO
  Team}}]{Rauer2016}
{Rauer}, H., {Aerts}, C., {Cabrera}, J., \& {PLATO Team}. 2016, Astronomische
  Nachrichten, 337, 961

\bibitem[{{Ricker} {et~al.}(2016){Ricker}, {Vanderspek}, {Winn}, {Seager},
  {Berta-Thompson}, {Levine}, {Villasenor}, {Latham}, {Charbonneau}, {Holman},
  {Johnson}, {Sasselov}, {Szentgyorgyi}, {Torres}, {Bakos}, {Brown},
  {Christensen-Dalsgaard}, {Kjeldsen}, {Clampin}, {Rinehart}, {Deming}, {Doty},
  {Dunham}, {Ida}, {Kawai}, {Sato}, {Jenkins}, {Lissauer}, {Jernigan},
  {Kaltenegger}, {Laughlin}, {Lin}, {McCullough}, {Narita}, {Pepper},
  {Stassun}, \& {Udry}}]{Ricker2016}
{Ricker}, G.~R., {Vanderspek}, R., {Winn}, J., {et~al.} 2016, in \procspie,
  Vol. 9904, Space Telescopes and Instrumentation 2016: Optical, Infrared, and
  Millimeter Wave, 99042B

\bibitem[{Rothman {et~al.}(2010)Rothman, Gordon, Barber, Dothe, Gamache,
  Goldman, Perevalov, Tashkun, \& Tennyson}]{Rothman2010}
Rothman, L., Gordon, I., Barber, R., {et~al.} 2010, Journal of Quantitative
  Spectroscopy and Radiative Transfer, 111, 2139

\bibitem[{Rothman {et~al.}(2013)Rothman, Gordon, Babikov, Barbe, Benner,
  Bernath, Birk, Bizzocchi, Boudon, Brown, {et~al.}}]{Rothman2012}
Rothman, L.~S., Gordon, I.~E., Babikov, Y., {et~al.} 2013, Journal of
  Quantitative Spectroscopy and Radiative Transfer, 130, 4

\bibitem[{Schaefer \& Fegley(2009)}]{Schaefer2009}
Schaefer, L., \& Fegley, B. 2009, The Astrophysical Journal Letters, 703, L113

\bibitem[{{Shkolnik} {et~al.}(2008){Shkolnik}, {Bohlender}, {Walker}, \&
  {Collier Cameron}}]{Shkolnik2008}
{Shkolnik}, E., {Bohlender}, D.~A., {Walker}, G.~A.~H., \& {Collier Cameron},
  A. 2008, \apj, 676, 628

\bibitem[{{Shporer}(2017)}]{Shporer2017}
{Shporer}, A. 2017, \pasp, 129, 072001

\bibitem[{{Shporer} \& {Hu}(2015)}]{ShporerHu2015}
{Shporer}, A., \& {Hu}, R. 2015, \aj, 150, 112

\bibitem[{{Stevenson} {et~al.}(2014){Stevenson}, {D{\'e}sert}, {Line}, {Bean},
  {Fortney}, {Showman}, {Kataria}, {Kreidberg}, {McCullough}, {Henry},
  {Charbonneau}, {Burrows}, {Seager}, {Madhusudhan}, {Williamson}, \&
  {Homeier}}]{Stevenson2014}
{Stevenson}, K.~B., {D{\'e}sert}, J.-M., {Line}, M.~R., {et~al.} 2014, Science,
  346, 838

\bibitem[{{Tsiaras} {et~al.}(2016){Tsiaras}, {Rocchetto}, {Waldmann}, {Venot},
  {Varley}, {Morello}, {Damiano}, {Tinetti}, {Barton}, {Yurchenko}, \&
  {Tennyson}}]{Tsiaras2016}
{Tsiaras}, A., {Rocchetto}, M., {Waldmann}, I.~P., {et~al.} 2016, \apj, 820, 99

\bibitem[{{von Braun} {et~al.}(2011){von Braun}, {Boyajian}, {ten Brummelaar},
  {Kane}, {van Belle}, {Ciardi}, {Raymond}, {L{\'o}pez-Morales}, {McAlister},
  {Schaefer}, {Ridgway}, {Sturmann}, {Sturmann}, {White}, {Turner},
  {Farrington}, \& {Goldfinger}}]{vonBraun2011}
{von Braun}, K., {Boyajian}, T.~S., {ten Brummelaar}, T.~A., {et~al.} 2011,
  \apj, 740, 49

\bibitem[{{Webber} {et~al.}(2015){Webber}, {Lewis}, {Marley}, {Morley},
  {Fortney}, \& {Cahoy}}]{Webber2015}
{Webber}, M.~W., {Lewis}, N.~K., {Marley}, M., {et~al.} 2015, \apj, 804, 94

\bibitem[{{Winn} {et~al.}(2011){Winn}, {Matthews}, {Dawson}, {Fabrycky},
  {Holman}, {Kallinger}, {Kuschnig}, {Sasselov}, {Dragomir}, {Guenther},
  {Moffat}, {Rowe}, {Rucinski}, \& {Weiss}}]{Winn2011}
{Winn}, J.~N., {Matthews}, J.~M., {Dawson}, R.~I., {et~al.} 2011, \apjl, 737,
  L18

\bibitem[{{Wong} {et~al.}(2016){Wong}, {Knutson}, {Kataria}, {Lewis},
  {Burrows}, {Fortney}, {Schwartz}, {Shporer}, {Agol}, {Cowan}, {Deming},
  {D{\'e}sert}, {Fulton}, {Howard}, {Langton}, {Laughlin}, {Showman}, \&
  {Todorov}}]{Wong2016}
{Wong}, I., {Knutson}, H.~A., {Kataria}, T., {et~al.} 2016, \apj, 823, 122

\bibitem[{{Zhang} \& {Showman}(2017)}]{ZhangShowman2017}
{Zhang}, X., \& {Showman}, A.~P. 2017, \apj, 836, 73

\end{thebibliography}
\end{document}